# Disentangling History and Propagation Dependencies in Cross-Subject Knee Contact Stress Prediction Using a Shared MeshGraphNet Backbone

Zhengye Pan[a,b], Jianwei Zuo[a,b], Jiajia Luo[*][a,b]

[a] Biomedical Engineering Department, Institute of Advanced Clinical Medicine, Peking University, Beijing, China

[b] Institute of Medical Technology, Peking University Health Science Center, Peking University, Beijing, China

[*] Corresponding author

**Abstract**

**Background:**

Subject-specific finite element analysis accurately characterizes knee joint mechanics but is computationally expensive. Deep surrogate models provide a near-real-time alternative, yet their generalization across subjects under limited pose and load inputs remains unclear. It remains unclear whether the dominant source of prediction uncertainty arises from temporal history dependence or spatial propagation dependence.

**Methods:**

To disentangle these factors, we employed a shared MeshGraphNet (MGN) backbone with a fixed mesh topology. A dataset of running trials from nine subjects was constructed using an OpenSim-FEBio workflow. We developed four model variants to isolate specific dependencies: (1) a baseline MGN; (2) CT-MGN, incorporating a Control Transformer to encode short-horizon history; (3) MsgModMGN, applying state-conditioned modulation to message passing for adaptive propagation; and (4) CT-MsgModMGN, combining both mechanisms. Models were evaluated using a rigorous grouped 3-fold cross-validation on unseen subjects.

**Results:**

The models incorporating history encoding (CT-MGN and CT-MsgModMGN) significantly outperformed the baseline MGN and MsgModMGN in global accuracy (lower RMSE) and spatial consistency (higher Dice/IoU). Crucially, the CT module effectively mitigated the "peak-shaving" defect common in deep surrogates, significantly reducing peak stress prediction errors ($RE_{\max}$). In contrast, the spatial propagation modulation (MsgMod) alone yielded no significant improvement over the baseline, and combining it with CT provided no additional benefit.

**Conclusion:**

Temporal history dependence, rather than spatial propagation modulation, is the primary driver of prediction uncertainty in cross-subject knee contact mechanics. Explicitly encoding short-horizon driver sequences enables the surrogate model to recover implicit phase information, thereby achieving superior fidelity in peak-stress capture and high-risk-region localization compared to purely state-based approaches.

# 1. Introduction

As a central hub of the lower-limb kinetic chain, the knee exhibits relatively limited osseous constraint; its dynamic stability therefore relies heavily on neuromuscular control and the coordinated restraint of soft tissues such as the ligaments and menisci [1,2]. Under conditions such as fatigue and high mechanical demand, the joint can readily shift into non-physiological loading, leading to aberrant contact stress distributions and an elevated risk of acute injury and osteoarthritis [3,4]. Accordingly, accurately quantifying the intra-articular mechanical environment is of substantial value for knee injury prevention and rehabilitation. Subject-specific knee finite element analysis (FEA) can faithfully characterize internal joint mechanics and, by coupling dynamic boundary conditions derived from multibody musculoskeletal dynamics, reconstruct the complex loading environment during movement with high agreement to experimental references in both contact magnitude and spatial distribution [5,6]. However, FEA typically requires labor-intensive, difficult-to-automate pre-processing and incurs considerable computational cost due to nonlinear iterations, limiting rapid reuse across individuals and postures and thereby constraining its practical efficiency in clinical settings.

Many studies have introduced deep surrogate models that learn nonlinear mappings between observable input drivers and tissue-scale mechanical responses, enabling near–real-time prediction of stress/strain fields at substantially reduced computational cost and offering a practical path to scale finite element analysis (FEA) to large cohorts and diverse conditions [7,8]. For example, Perrone et al. [9] developed a "gait-to-contact" deep learning framework that reduced high-fidelity FE wear prediction from days of computation to inference in minutes while achieving high structural similarity (SSIM > 0.88). Liu et al. [10] built a supervised surrogate that bypassed labor-intensive FE pre-processing and repeated solves, enabling rapid prediction of soft-tissue mechanical responses under varying degrees of meniscal extrusion. However, progress in this area has largely been driven by minimizing global error metrics, with limited effort devoted to disentangling the dominant sources of cross-subject uncertainty[11-13]. When inputs are restricted to a small set of drivers, such as joint pose and net load, a key unresolved question is whether uncertainty in cross-subject stress-field prediction is governed primarily by temporal history dependence or by spatial propagation dependence. If stress responses depend strongly on the recent loading trajectory and phase, short-horizon driver sequences must be encoded to recover implicit phase/transition states and reduce path ambiguities that cannot be resolved from a single frame alone [14]. Conversely, if errors are dominated by state-dependent changes in nonlocal load transfer across coupled tissues, a state-sensitive propagation mechanism is needed on a fixed anatomical topology. Such a mechanism would allow message passing to adaptively modulate the effective transfer gain (i.e., coupling strength) [15,16], improving fidelity in peaks and the spatial structure of high-risk regions. These mechanisms may be substitutable or complementary; yet few studies have compared their relative contributions under a unified input setting and matched model capacity.

To address this gap, we adopt a shared MGN backbone with a fixed mesh topology and identical inputs, and construct three targeted variants: CT-MGN, which augments MGN with a Control Transformer (CT) to encode short-horizon driver sequences and provide temporal context; MsgModMGN, which applies state-conditioned modulation to message passing to enable adaptive propagation; and CT-MsgModMGN, which combines both components. Using the same training pipeline and matched model capacity, we systematically compare all four models in terms of peak error and the reconstruction consistency of high-risk regions, thereby quantifying the relative contributions of propagation dependence and history dependence to cross-subject prediction of knee mechanical responses and providing guidance for designing clinically generalizable surrogate models of joint mechanics.

## 2 Methods

### 2.1 Dataset construction

2.1.1 Participant recruitment and data acquisition

Nine healthy adult male participants were enrolled (height: $176.41 \pm 4.60$ cm; body mass: $73.83 \pm 6.90$ kg; age: $23.51 \pm 4.73$ years). Inclusion criteria were no history of knee pain or prior knee surgery and no apparent lower-limb functional impairments. Before data collection, the study objectives and procedures were fully explained to all participants, and written informed consent was obtained.

Kinematic data were collected using a Vicon optical motion-capture system (200 Hz, V5,

Oxford, UK), and ground reaction forces were synchronously recorded with a Kistler 3D force plate (1000 Hz, 9286AA, Winterthur, Switzerland). The reflective marker set followed the protocol described by Pan et al. [17]. To control running speed, a single-beam infrared timing gate was positioned 1 m in front of the force plate center to monitor the participant's instantaneous speed during each trial. After a standardized warm-up, each participant performed five running trials at a target speed of 2.5 m/s; the three trials with full-foot contact on the force plate closest to the target speed were selected for subsequent analyses. The stance phase was identified using a 20 N threshold on the vertical ground reaction force (vGRF) [18]: initial contact was defined as vGRF > 20 N, toe-off as vGRF < 20 N, and the interval between these events was considered the stance phase.

2.1.2 Data processing

Both the marker trajectories and the GRF were filtered using a fourth-order, zero-lag (bidirectional) Butterworth low-pass filter with a cutoff frequency of 10 Hz [19]. Subsequently, inverse kinematics and joint reaction analyses were performed in OpenSim using the full-body musculoskeletal model proposed by Rajagopal et al. [20], in which the knee joint was modeled with three rotational degrees of freedom: flexion–extension, internal–external rotation, and adduction–abduction. Knee kinematics were represented as XYZ Euler angles expressed in the tibial coordinate system, and the joint reaction force vector was resolved in the same joint coordinate system. Together, these quantities served as boundary conditions for the subsequent finite element analyses.

2.1.3 Finite element simulation and dataset generation

To mitigate the influence of inter-subject anatomical variability on model evaluation, we used the generic finite element mesh as a unified anatomical and topological reference [21]. The knee pose computed in OpenSim was prescribed as kinematic constraints, and the joint reaction force was applied as an equivalent external load at a distal femoral reference node to reproduce boundary loading consistent with the measured motion. Kinematics-driven quasi-static analyses were conducted in FEBio with a time-step size of 0.01. For material modeling, to capture fiber-reinforcement–induced anisotropy, the menisci and ligaments were modeled using a transversely isotropic Mooney–Rivlin hyperelastic formulation (with initial ligament tension introduced via pre-strain). Articular cartilage was modeled as an isotropic Mooney–Rivlin hyperelastic material.

Each time step of the quasi-static analysis was treated as an independent mechanical state and packaged into a graph sample $G = (V, E)$, where $V$ and $E$ denote the node and edge sets defined by the finite element mesh topology (with edges determined by mesh connectivity). Finite elements in the knee model were treated as graph nodes, and the element centroid coordinates served as the geometric representation of each node. For a given node $i$, the input feature vector $\mathbf{x}_i \in \mathbb{R}^{10}$ was formed by concatenating: (1) the centroid coordinates in the reference configuration, providing a geometric prior; and (2) a global driver vector, consisting of the joint reaction forces and joint pose encoded using sine/cosine terms, representing the time-varying load–pose condition. This global vector was broadcast to all nodes at each time step. The target label was the element-wise von Mises stress, which was first transformed using $\log(1 + x)$ and then Z-score normalized using the dataset mean and standard deviation.

**2.2 Deep surrogate models**

2.2.1 Network architecture

Finite element meshes are topologically isomorphic to message-passing graphs, and intra-articular load transfer and stress redistribution can be cast as nonlocal information propagation along mesh connectivity [22]. Accordingly, we adopted MGN as the shared backbone. Building on this backbone, we introduced (i) a short-horizon history encoder (Control Transformer; CT) and (ii) state-conditioned message modulation (MsgMod), yielding two variants: CT-MGN and MsgModMGN. We further combined both components to form CT-MsgModMGN, which integrates history encoding with propagation modulation.

The MGN backbone comprises a node/edge encoder, a K-step message-passing cap, and a node decoder. In the encoding stage, node features and edge geometric features are mapped into latent representations:

$$\mathbf{h}_i^{(0)} = f_v^{enc}(\mathbf{x}_i) \tag{1}$$

$$\mathbf{e}_{ij}^{(0)} = f_e^{enc}(\Delta \mathbf{p}_{ij}) \tag{2}$$

Where $\mathbf{x}_i$ is the node input, $\Delta \mathbf{p}_{ij}$ is the relative centroid displacement, and $f_v^{enc}$ and $f_e^{enc}$ are MLP encoders producing latent embeddings $\mathbf{h}_i^{(0)}$ and $\mathbf{e}_{ij}^{(0)}$. The processor performs $K$ iterations of message passing over a fixed topology. At each iteration, edge messages are first computed from the states of the incident nodes and the edge itself to update edge representations, after which incoming edge messages are aggregated to update node representations. The update can be summarized as:

$$h_i^{(k+1)} = h_i^{(k)} + \phi_v\left([h_i^{(k)}, \text{Agg}_{j \in \mathcal{N}(i)} \phi_e([h_j^{(k)}, h_i^{(k)}, e_{ji}^{(k)}])]\right) \qquad (3)$$

where Agg(·) denotes mean aggregation over the neighborhood, and $\phi_e(\cdot)$ and $\phi_v(\cdot)$ are multilayer perceptrons. To stabilize iterative propagation, residual connections and normalization are applied, and the same set of parameters is shared across message-passing steps. Finally, the decoder maps the final node representation $\mathbf{h}_i^{(K)}$ to a scalar stress prediction $\hat{y}_i$, enabling full-field reconstruction of joint contact stress.

Because instantaneous pose and net load do not uniquely determine the contact state, the underlying state information is often inferred from the recent temporal trajectory. We therefore augment the MGN backbone with a CT-based history encoder. Specifically, at each time step *t*, in addition to the current driver vector $\mathbf{d}_t$, we construct a length-*L* short-horizon driver sequence $\mathbf{D}_t = [\mathbf{d}_{t-L+1}, \ldots, \mathbf{d}_t]$. The CT module models this sequence and outputs a graph-level historical context vector $\mathbf{c}_t$ as a compact summary of the recent history:

$$c_t = \text{CT}(D_t) \qquad (4)$$

The resulting $\mathbf{c}_t$ is then injected into a node-wise conditioning pathway to generate FiLM parameters and modulate the node hidden representations. Accordingly, the core idea of CT-MGN is to extend single-frame drivers to a short-horizon driver trajectory.

In addition, the pathways and coupling strength of intra-articular load transfer can be reconfigured during movement as a function of joint pose and net-load state. To explicitly capture this propagation dependence, we introduce a state-conditioned message-modulation mechanism on top of the MGN backbone, forming MsgModMGN. Specifically, at each message-passing step, we first aggregate the current driver state into a graph-level context vector $\mathbf{c}_t$. A lightweight modulation network then maps $\mathbf{c}_t$ to channel-wise scaling coefficients $\mathbf{s}(\mathbf{c}_t)$, which are broadcast to edge messages to apply multiplicative modulation:

$$\widetilde{\mathbf{m}}_{ij}^{(k)} = s(\mathbf{c}_t) \odot \mathbf{m}_{ij}^{(k)} \qquad (5)$$

where $\mathbf{m}_{ij}^{(k)}$ denotes the edge message produced at step *k* by the edge update network. This mechanism is equivalent to learning a state-conditioned propagation operator on a fixed topology [23]. Without changing the input variable set or the capacity of the backbone network, it allows message passing to automatically adjust the effective transfer "gain" according to the current driver state.

Building on these components, CT-MsgModMGN uses the historical context $\mathbf{c}_t$ extracted by the CT module from the short-horizon driver sequence for two purposes: (i) node-wise FiLM conditioning to recover implicit loading phase/state information, and (ii) state-conditioned scaling within message passing to modulate propagation gain adaptively. In this way, CT-MsgModMGN provides a unified treatment of history dependence and propagation dependence.

2.2.2 Training and inference

All four models were supervised using element-wise von Mises stress. The network output was the scalar stress value after log(1 + x) transformation and Z-score normalization. To prevent invalid elements from degrading regression stability, errors were computed only over nodes marked as valid by a mask during training and validation. Accordingly, a masked mean squared error (masked MSE) was used as the training objective:

$$\mathcal{L}(\theta) = \frac{1}{\sum_i m_i} \sum_i m_i \parallel \hat{y}_i - y_i \parallel_2^2 \tag{6}$$

Where $m_i \in \{0,1\}$ is the node mask, and $y_i$ and $\hat{y}_i$ denote the normalized stress label and prediction, respectively. During evaluation, model outputs were mapped back to the physical scale by applying the inverse transformations used in training. All models were optimized using the same training pipeline and hyperparameter strategy to ensure fair comparison. Parameter updates were performed with the AdamW optimizer. A reduce-on-plateau learning-rate scheduler was applied based on the validation loss, and early stopping was used to mitigate overfitting. Given the graph size and GPU memory constraints, we trained with a small batch size and applied gradient-norm clipping to stabilize convergence. Automatic mixed precision was enabled for all models to improve training throughput. To control model capacity, all four models used the same backbone hidden dimension and message-passing depth, and the processor shared parameters across propagation steps. Hyperparameters were selected via a lightweight ablation study, and a compact configuration (Dhidden=48, K=3) was found to offer a favorable trade-off between efficiency and accuracy. All models were trained on two NVIDIA TITAN GPUs (batch size = 1; 120 epochs).

We conducted a grouped 3-fold cross-validation by partitioning the nine participants into three subject-level folds (Fold 1: P1–P3; Fold 2: P4–P6; Fold 3: P7–P9). In each fold, models were trained on data from two participant groups and evaluated on the remaining group of entirely unseen subjects. This subject-wise split prevents inter-subject data leakage by design. During inference, MGN and MsgModMGN predicted the stress field independently at each time step. In contrast, CT-MGN and CT-MsgModMGN additionally consumed a short-horizon driver sequence of length $L = 8$ to generate a historical context vector, which was then used to regress the stress field at the current time step.

2.2.3 Evaluation metrics

We evaluated the stress-field reconstruction performance of the four models from three perspectives: global error, peak fidelity, and spatial consistency. For global error, we quantified overall prediction deviation using the root-mean-square error (RMSE) and mean absolute error (MAE), and further reported the normalized RMSE (nRMSE) to remove scale effects. Let $N$ denote the number of nodes at time step $t$, where $y_i$ and $\hat{y}_i$ are the finite element ground-truth stress and the model prediction at node $i$, respectively. The metrics were computed as follows:

$$\text{RMSE} = \sqrt{\frac{1}{N}\sum_{i=1}^{N}(\hat{y}_i - y_i)^2} \tag{7}$$

$$\text{MAE} = \frac{1}{N}\sum_{i=1}^{N}|\hat{y}_i - y_i| \tag{8}$$

$$\text{nRMSE} = \frac{\text{RMSE}}{\max_j(y_j)} \tag{9}$$

nRMSE was normalized by the peak ground-truth stress at the current time step, $\max(y)$, thereby directly reflecting the prediction error as a fraction of the instantaneous maximum loading level. Because injury-related mechanisms are often dominated by localized high-stress concentrations, global metrics such as RMSE/MAE may mask underestimation of extreme values. We therefore defined the peak relative error ($RE_{max}$) and the 95th-percentile relative error ($RE_{P95}$) to specifically assess whether the models exhibit the "peak-shaving" behavior commonly observed in deep surrogate models:

$$RE_{max} = \frac{|max_i(\hat{y}_i) - max_i(y_i)|}{max_i(y_i)} \tag{10}$$

$$RE_{P95} = \frac{|P_{95}(\hat{y}) - P_{95}(y)|}{P_{95}(y)} \tag{11}$$

These two metrics are well-suited to capturing prediction bias in the most critical, high-stress regions of the meniscus. In addition, we computed the node-wise Pearson correlation coefficient ($r$) and used the Dice coefficient and intersection over union (IoU) to quantify the spatial overlap of high-risk regions:

$$r = \frac{\Sigma(\hat{y}_i-\bar{\hat{y}})(y_i-\bar{y})}{\sqrt{\Sigma(\hat{y}_i-\bar{\hat{y}})^2}\sqrt{\Sigma(y_i-\bar{y})^2}} \tag{12}$$

$$\text{Dice} = \frac{2|S_{true}\cap S_{pred}|}{|S_{true}|+|S_{pred}|} \tag{13}$$

$$\text{IoU} = \frac{|S_{\text{pred}}\cap S_{\text{true}}|}{|S_{\text{pred}}\cup S_{\text{true}}|} \tag{14}$$

where the set $S_t^{true} = \{\, i \mid y_{t,i} > P_{95}(y_t)\,\}$ denotes the high-risk region, i.e., nodes whose ground-truth stress exceeds the 95th-percentile threshold. To quantify hotspot localization, we further introduced the hotspot distance, which measures the spatial offset between the ground-truth and predicted high-risk regions. Let $p_i$ denote the centroid coordinate of node (or element) $i$; the geometric centers of the ground-truth and predicted high-risk regions are defined as:

$$\mathbf{c}_t^{true} = \frac{1}{|S_t^{true}|}\Sigma_{i\in S_t^{true}}\,\mathbf{p}_i \tag{15}$$

$$\mathbf{c}_t^{pred} = \frac{1}{|S_t^{pred}|}\Sigma_{i\in S_t^{pred}}\,\mathbf{p}_i \tag{16}$$

and the hotspot distance is defined as $d_t^{hot}$, which quantifies the spatial displacement between the two hotspot regions.

$$d_t^{hot} =\parallel \mathbf{c}_t^{true} - \mathbf{c}_t^{pred} \parallel_2 \tag{17}$$

**2.3 Statistical analysis**

To compare the four models across evaluation metrics, we applied the Friedman nonparametric test to paired repeated-measures data to assess overall differences among models. When the omnibus test was significant, we performed pairwise Wilcoxon signed-rank tests for post hoc comparisons, with multiplicity controlled using the Holm–Bonferroni procedure. All tests were two-sided with a significance level of $\alpha = 0.05$. Metrics are reported as mean ± standard deviation (Mean ± SD), and all statistical analyses were performed in Python.

# 3 Results

**3.1 Full-field stress reconstruction accuracy**

Overall full-field reconstruction errors for the four models are summarized in Table 1. CT-MGN and CT-MsgModMGN achieved significantly lower RMSE and MAE than MGN and MsgModMGN ($p < 0.05$). No significant differences in RMSE or MAE were observed between MsgModMGN and MGN. In addition, CT-MGN yielded significantly lower overall error than CT-MsgModMGN ($p < 0.05$).

Table 1. Comparison of full-field nodal stress prediction errors across the four models (mean ± SD).

|  | MGN | MsgModMGN | CT-MGN | CT-MsgModMGN |
|---|---|---|---|---|
| RMSE | 0.60 ± 0.15 | 0.56 ± 0.11 | 0.37 ± 0.08* | 0.42 ± 0.10 |
| MAE | 0.25 ± 0.06 | 0.23 ± 0.05 | 0.12 ± 0.03* | 0.15 ± 0.04 |

**Note:** * indicates statistical significance compared to all other models.

Consistency and normalized error for full-field stress reconstruction are shown in Figure 1. Consistent with the patterns in Table 1, CT-MGN and CT-MsgModMGN achieved significantly higher Pearson correlation ($r$) than MGN and MsgModMGN ($p < 0.05$), along with significantly lower nRMSE ($p < 0.05$). In contrast, MsgModMGN did not differ significantly from MGN in either $r$ or nRMSE. Moreover, CT-MGN remained significantly better than CT-MsgModMGN in both $r$ and nRMSE ($p < 0.05$).

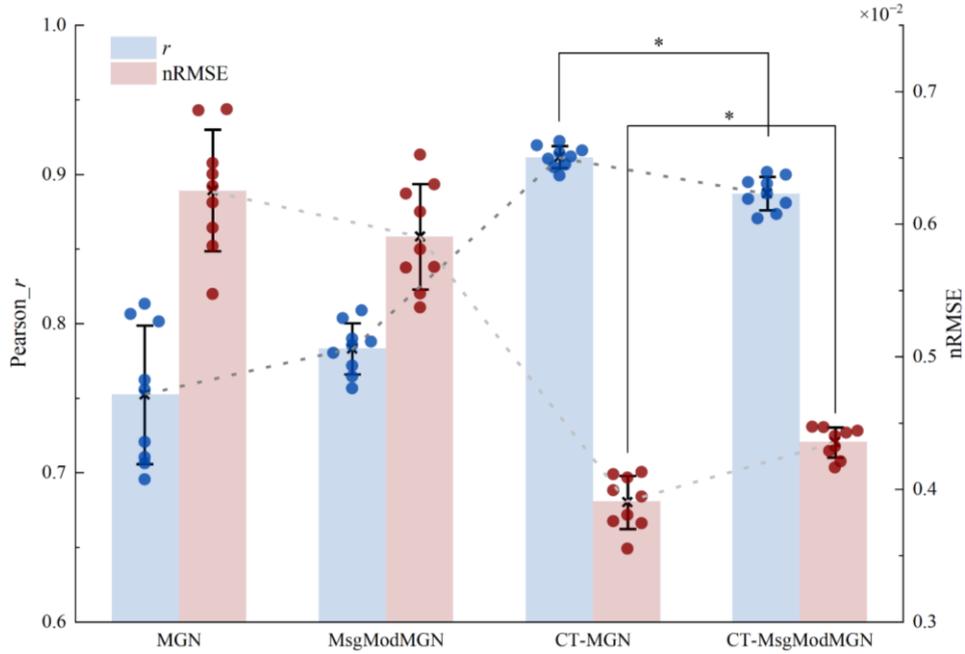

Figure 1. Full-field stress reconstruction consistency and normalized error across the four models. * indicates a significant difference between CT-MGN and CT-MsgModMGN. Solid circles represent individual subjects and are horizontally jittered around the corresponding bar center for clarity; the x-position has no statistical meaning. nRMSE and $r$ are shown in red and blue, respectively.

### 3.2 Accuracy in reconstructing peak and high-quantile stress

As shown in Figure 2, for both the peak relative error ($RE_{max}$) and the 95th-percentile relative error ($RE_{P95}$), CT-MGN and CT-MsgModMGN performed significantly better than MGN and MsgModMGN ($p$ = 0.0234). No significant differences were observed between CT-MGN and CT-MsgModMGN, and likewise between MGN and MsgModMGN.

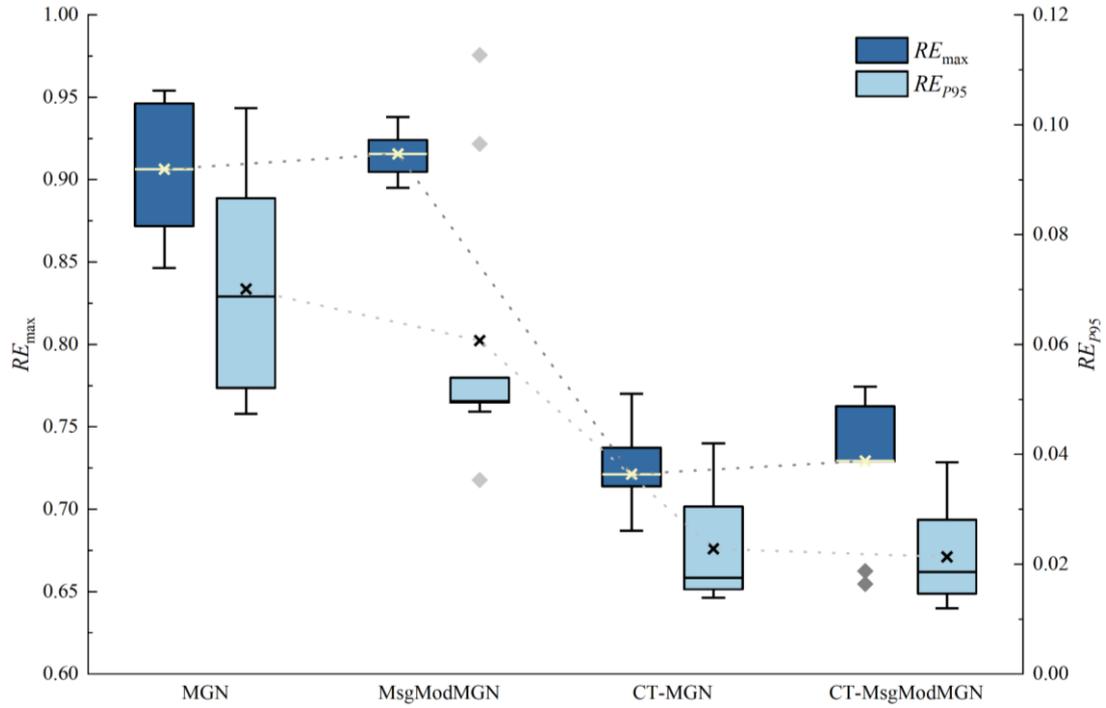

Figure 2. Comparison of peak and high-quantile stress reconstruction errors across the four models (diamond markers indicate outliers).

### 3.3 High-risk region consistency and temporal evolution of high-quantile error

As shown in Figure 3, CT-MGN and CT-MsgModMGN achieved significantly higher consistency in reconstructing high-risk regions than MGN and MsgModMGN ($p$ < 0.05). No significant differences in Dice/IoU were observed between MGN and MsgModMGN. Similarly, CT-MGN and CT-MsgModMGN did not differ significantly in Dice/IoU, indicating comparable spatial overlap of high-risk regions between the two CT-based models.

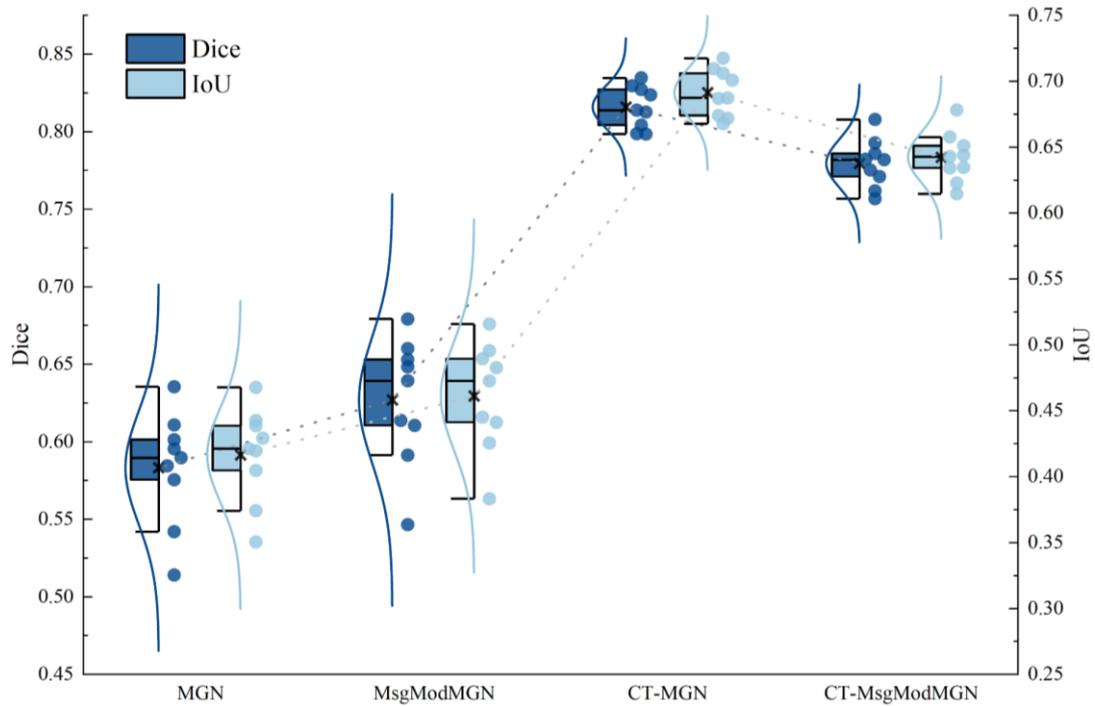

Figure 3. Comparison of the consistency of high-risk region reconstruction across the four models. Solid dots represent individual subjects; the overlaid normal distribution curves summarize the across-subject distribution; × denotes the mean; the box indicates the interquartile range, and the line inside the box marks the median.

The temporal evolution of $RE_{P95}$ over 0-100% of stance is shown in Figure 4. Overall, CT-MGN and CT-MsgModMGN maintained lower $RE_{P95}$ throughout the stance phase, whereas MGN and MsgModMGN exhibited higher $RE_{P95}$ with greater variability. Notably, during mid-to-late stance (approximately 50-75%), the non-CT models exhibited a more pronounced increase in REP95 and greater uncertainty. In contrast, the CT-based models remained relatively stable over the same interval. This temporal pattern is consistent with the high-risk region consistency results shown in Figure 3.

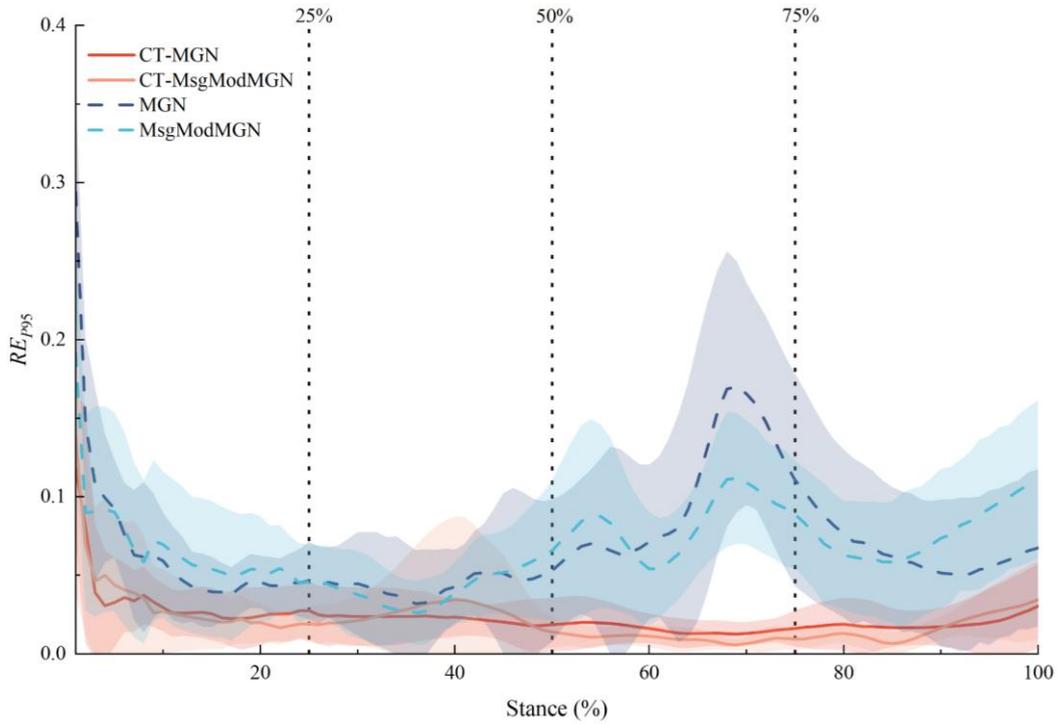

Figure 4. Temporal evolution of the high-quantile error $RE_{P95}$ over the stance phase. Solid/dashed lines show the mean across subjects, and the shaded bands indicate ± SD. Vertical dotted lines mark 25%, 50%, and 75% of stance.

## 3.4 Hotspot localization performance

As shown in Figure 5, the four models exhibited a stratified pattern in hotspot offset distance that mirrors the results for high-risk region consistency. CT-MGN and CT-MsgModMGN maintained lower hotspot offset distances throughout stance with a narrower variability band, indicating more stable localization of the high-risk region. In contrast, MGN and MsgModMGN showed larger hotspot offsets and greater uncertainty, with more pronounced deviations and fluctuations during early and mid-stance. Notably, around mid-stance (approximately 50%), the non-CT models exhibited a clearer increase in offset distance. In contrast, the CT-based models remained relatively steady, demonstrating more reliable hotspot localization during the critical load-bearing phase.

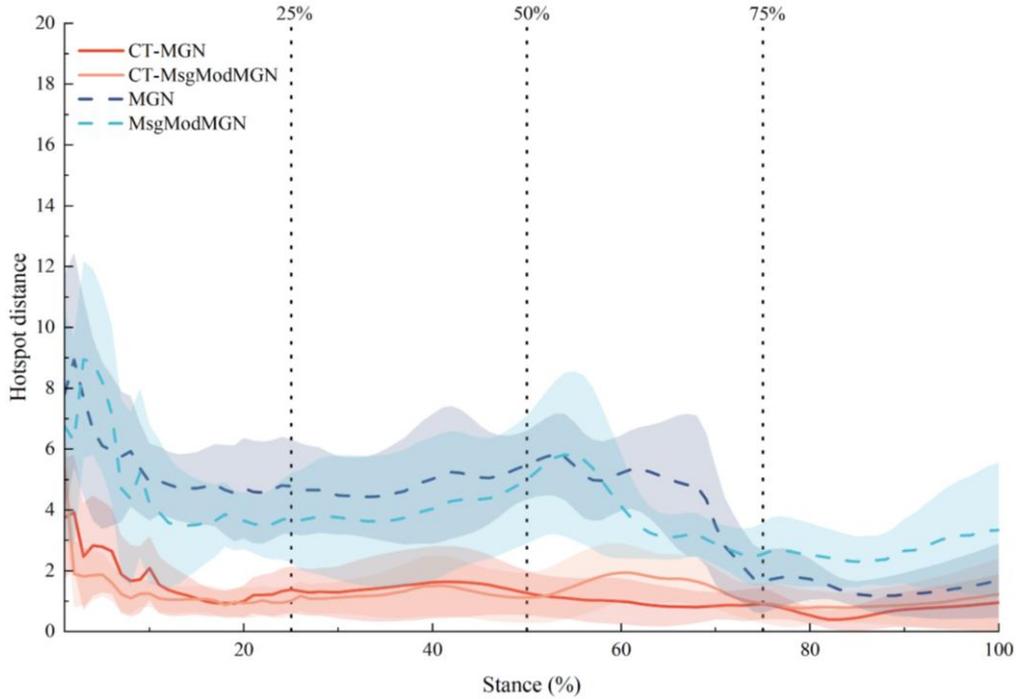

Figure 5. Temporal evolution of hotspot localization error over the stance phase. Solid/dashed lines show the mean across subjects, and the shaded bands indicate ± SD. Vertical dotted lines mark 25%, 50%, and 75% of stance.

## 4 Discussion

This study investigates the dominant sources of uncertainty in cross-subject prediction of knee stress fields. Under a unified input setting, fixed-mesh topology, and matched training/model-capacity configuration, we incorporate a short-horizon history-encoding CT module and a driver-state–conditioned message-modulation mechanism, and systematically compare the effects of temporal history dependence and spatial propagation dependence on cross-subject stress-field reconstruction performance.

The results show that the CT module yields a pronounced improvement in prediction accuracy: CT-MGN and CT-MsgModMGN achieved significantly lower RMSE and MAE than the models without history encoding (Table 1). Consistently, the CT-based models produced stress distributions that more closely matched the finite element ground truth, exhibiting significantly higher node-wise correlation ($r$) and significantly lower normalized error (nRMSE) (Fig. 1). These findings indicate that short-horizon sequences provide phase/transition-state information that substantially strengthens the model's ability to represent instantaneous load variations, thereby reducing biases arising from the inability of single-frame inputs to disambiguate differences in loading trajectories. For example, during phases of rapid stress variation—such as mid-to-late stance—the CT-based models can leverage recent loading history to prevent error accumulation and the expansion of temporal fluctuations (Fig. 4), thereby maintaining stable prediction accuracy across the gait cycle. In contrast, introducing MsgMod alone did not lead to a significant improvement in overall performance, suggesting that merely adjusting message-passing gain based on the instantaneous driver state is insufficient to reduce global error substantially. Collectively, these results imply that the dominant uncertainty in stress-field prediction may stem more from temporal history dependence than from pose-induced differences in spatial propagation patterns. In other words, missing short-horizon context may be a primary driver of the accuracy bottleneck in conventional deep surrogates, and incorporating temporal memory provides a viable route to alleviate this limitation.

With respect to peak stress and high-risk region reconstruction, the CT module also contributed substantially. CT-MGN and CT-MsgModMGN achieved significantly lower peak relative error and high-quantile error, markedly mitigating the "peak-shaving" tendency commonly observed in deep surrogate models (Fig. 2). This improvement reduced underestimation of extreme stresses. It brought predicted peak magnitudes closer to the finite element reference, which is critical for injury-related risk assessment. Moreover, relative to the baseline model, incorporating CT yielded large gains in the Dice coefficient and IoU for high-risk regions (Fig. 3), indicating more accurate identification of stress hotspots in vulnerable structures, such as the meniscus. In addition, the CT-based models exhibited smaller and less time-varying offsets between the geometric centers of the predicted and ground-truth hotspot regions, indicating more stable and reliable localization of high-risk areas (Fig. 5). By contrast, MsgMod alone conferred no apparent advantage on these metrics: adding MsgMod neither significantly reduced peak error nor improved spatial overlap of high-stress regions. This suggests that, in the current setting, the baseline MeshGraphNet already captures nonlocal load-transfer patterns on a fixed topology to some extent, and state-conditioned modulation of message passing alone is insufficient to materially improve peak reconstruction or hotspot identification. Notably, when CT and MsgMod were combined (CT-MsgModMGN), performance did not surpass that of CT-MGN; overall error was slightly higher (Table 1). This indicates that the two mechanisms are not fully complementary under our experimental conditions: the key information introduced by history encoding may already account for the dominant uncertainty, leaving limited room for MsgMod to provide additional benefit and potentially introducing a mild risk of overfitting due to the added parameters.

Although this study demonstrates the effectiveness of incorporating history encoding to improve knee stress prediction, the dataset is limited in size and diversity, which may restrict generalization to broader populations and a wider range of movement patterns. In addition, simplifying assumptions in both finite element modeling and graph-based representation should be acknowledged. Specifically, ground-truth stresses were obtained from quasi-static finite element analyses, which may partially neglect dynamic effects such as inertia and viscoelasticity. Future work could therefore incorporate higher-fidelity finite element modeling to better capture these dynamics.

## 5 Conclusion

Using MeshGraphNet as a shared backbone, this study quantitatively assessed the relative contributions of short-horizon history encoding and driver-state–conditioned message modulation to model performance, and found that the CT module provides substantially greater benefits than MsgMod. Incorporating short-horizon temporal context effectively compensates for limited sensitivity to dynamic loading trajectories, yielding clear improvements in overall accuracy, peak-stress fidelity, and high-stress region localization. In contrast, state-conditioned message passing had a limited impact under the relatively constrained range of pose/load variations considered here. These findings provide practical guidance for designing deep models for joint mechanics field prediction: prioritizing the acquisition of temporal dependencies, and considering finer-grained spatial propagation modulation only after the key temporal features have been robustly learned. In more complex and heterogeneous conditions, MsgMod may offer additional value, but realizing its potential will likely require richer data and more targeted modeling strategies.

## Funding Statement

This study was supported by the National Key R&D Program of China (grant no. 2023YFC2411201); Beijing Natural Science Foundation (grant no. L259081); NSFC General Program (grant no. 31870942); Peking University Clinical Medicine Plus X – Young Scholars Project (grant nos. PKU2020LCXQ017 and PKU2021LCXQ028); and PKU-Baidu Fund (grant no. 2020BD039).

## Declaration of Interest Statement

The authors have no conflicts of interest to declare.